\documentclass[letterpaper,draftcls,onecolumn,journal]{IEEEtran}

\input epsf
\usepackage{graphicx}
\usepackage{amsmath}   
\usepackage{siunitx}
\usepackage{color}
\usepackage{psfrag}
\usepackage[dvips]{pstricks}
\usepackage{url}
\usepackage{booktabs}
\usepackage{threeparttable}

\usepackage{endfloat}

\usepackage{cite}

\newcommand{\oldrev}[1]{#1}
\newcommand{\newrev}[1]{#1}

\begin{document}

\title{Generation of reference dc currents at \SI{1}{\nano\ampere} level with the capacitance-charging method}

%
\author{Luca Callegaro, Pier Paolo Capra, Vincenzo D'Elia and Flavio Galliana
\thanks{The authors are with the Electromagnetism Division of the Istituto Nazionale di Ricerca Metrologica (INRIM), strada delle Cacce 91, 10135 Torino, Italy, e-mail: \texttt{l.callegaro@inrim.it}}
\thanks{Manuscript received \today.}}


\maketitle

\begin{abstract}
The capacitance-charging method is a well-established and handy technique for the generation of dc current in the \SI{100}{\pico\ampere} range or lower. The method involves a capacitance standard and a sampling voltmeter, highly stable devices easy to calibrate, \newrev{and} it is robust and insensitive to the voltage burden of the instrument being calibrated. We propose here a range extender amplifier, which can be employed as a plug-in component in existing calibration setups, and allows the generation of currents in the \SI{1}{\nano\ampere} range. The extender has been employed in the INRIM setup and validated with two comparisons at \SI{100}{\pico\ampere} and \SI{1}{\nano\ampere} current level.  The calibration accuracy achieved on a top-class instrument is \num{5E-5} at is \SI{1}{\nano\ampere}. 
\end{abstract}

\IEEEoverridecommandlockouts
\begin{IEEEkeywords}
Metrology; current measurement; measurement techniques; calibration; measurement uncertainty; measurement standards.
\end{IEEEkeywords}

%
\IEEEpeerreviewmaketitle

\section{Introduction}

\IEEEPARstart{T}{he} accurate generation of low dc current values is of relevance for the calibration of low dc current meters (picoammeters, femtoammeters) and for research on electron-counting sources based on single-electron tunneling. \newrev{Methods based on Ohm's law,} where a current $I$ is generated by applying a dc voltage $V$ \newrev{to} a resistor $R$, becomes increasingly impractical for lower and lower currents. \newrev{The primary} reasons are that very high-valued resistance standards required have high voltage and temperature coefficients, long time constants, are difficult to calibrate; and that the burden voltage of the instrument under calibration is an important source of error.

Recently, several national metrology institutes have developed generators \newrev{\cite{Willenberg2003,Brom2005,Callegaro2007b,Fletcher2007,Iisakka2008,Willenberg2009,Bergsten2011}} based on the displacement current in a capacitor, called \emph{capacitance charging} method. A linear voltage ramp $v(t) = St$, having slope $S$, is applied to a capacitor \oldrev{C (having capacitance $C$)}, and a current $\displaystyle I = C \frac{\mathrm{d} v(t)}{\mathrm{d} t} = SC$ is generated. Appropriate capacitance standards to be employed in the method can be highly stable and can be accurately calibrated; \oldrev{$S$} can be calculated from readings of a sampling voltmeter. The method is not affected by the ammeter burden voltage, which affects $v(t)$ but not its slope $S$.

\newrev{The capacitor} must be a gas-dielectric or vacuum capacitance standard, since all solid-dielectric capacitors show the phenomenon of dielectric absorption \cite{Pease1982} which would severely limit the generator accuracy. Available commercial standard capacitors range from \SI{1}{\pico\farad} to \SI{10}{\nano\farad}\footnote{Values lower than \SI{1}{\pico\farad} can be achieved with the Zickner construction \cite{Zickner1930}.}. However, capacitances above \SI{1}{\nano\farad} are likely to \newrev{cause} oscillations in the input stage of the instrument being calibrated \cite{Pease2001} and therefore cannot be employed.

\newrev{The} maximum $\displaystyle S = \frac{2 V_\mathrm{max}}{T}$ is limited by the maximum voltage $V_\mathrm{max}$ available by the ramp generator and the ramp duration $T$, which should be adequate for measurement settling. 

In most published generators \cite{Brom2005,Callegaro2007b,Iisakka2008,Bergsten2011}, $T=$\SIrange{100}{200}{\second} and $V_\mathrm{max}=\SI{10}{\volt}$, which in turn limits the maximum available current to \SIrange{100}{200}{\pico\ampere} when $S=\SI{0.1}{\volt\per\second}$ and $C=\SI{1}{\nano\farad}$. In one implementation \cite{Fletcher2007} a maximum current of \SI{1}{\nano\ampere} has been reached at the expense of raising the current slope to $S=\SI{1}{\volt\per\second}$, hence reducing the available measurement time to \SI{20}{\second}.

In the following, we investigate a simple range extender amplifier, that can be connected to an existing ramp generator. It raises \newrev{the voltage to the} \SI{100}{\volt} range, and thus permits \newrev{the extension of} the maximum current range of the capacitance-charging method to \SI{1}{\nano\ampere} range. Calibration measurements performed on a top-level instrument are compared with those obtained with Ohm's law method.

\section{Measurement setup}
\begin{figure}[tb]
 	\includegraphics[width=\linewidth]{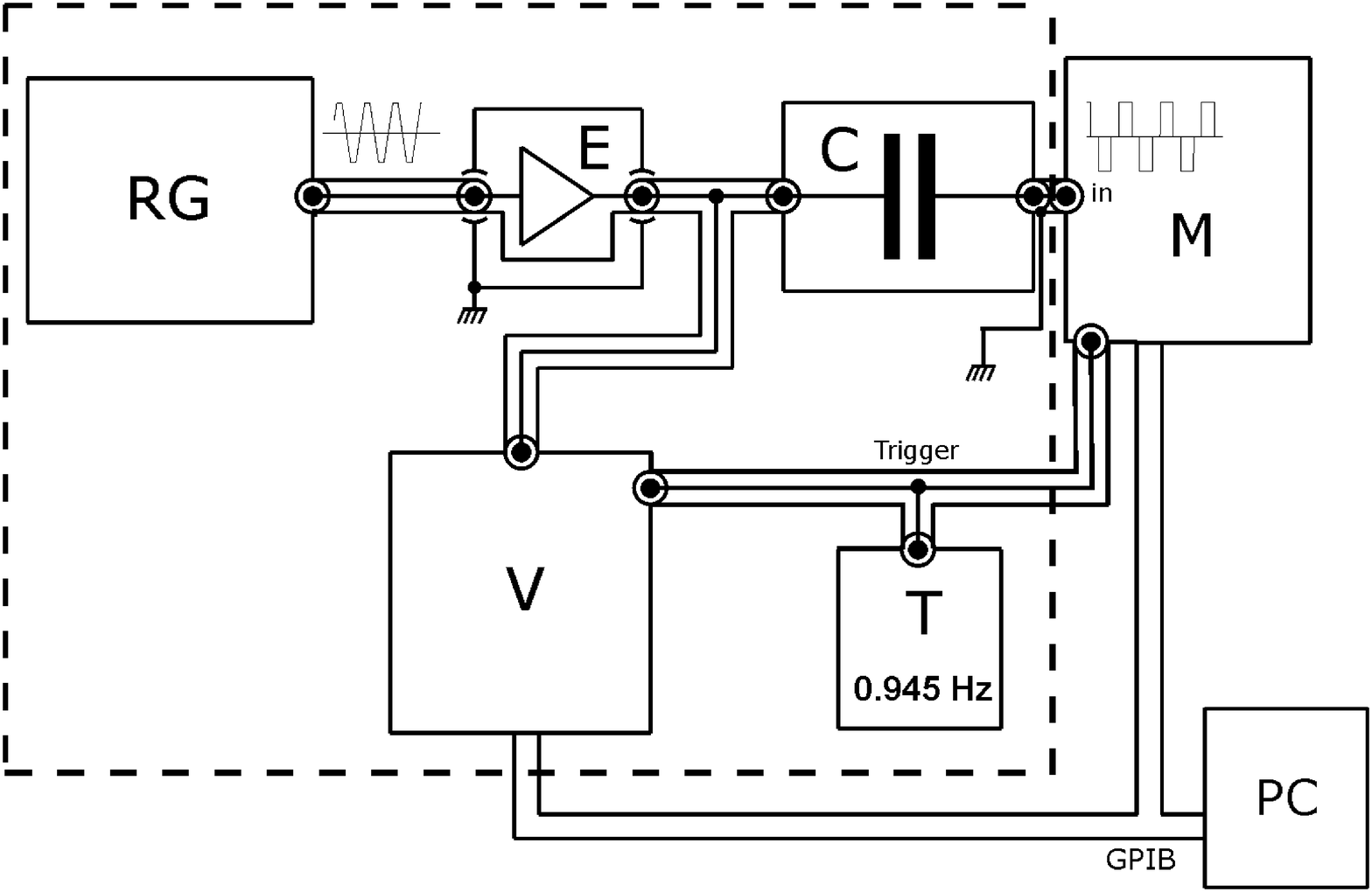}%
 	\caption{Block schematics of the measurement setup. RG is the ramp generator, E the range extender amplifier, \oldrev{C} the current-generating gas-dielectric capacitor, V the sampling voltmeter, T the timebase generator, M the meter being calibrated, PC the data acquisition computer. \label{fig:capsetup}}%
\end{figure}

The measurement setup is shown in Fig. \ref{fig:capsetup}. The range extender amplifier E has a $\times 10$ gain configuration, and is connected to the existing ramp generator RG  \cite{Callegaro2007b} (having $V_\mathrm{max} = \SI{10}{\volt}$\newrev{)}. The ramp voltage $v(t)$, sampled by the voltmeter V (Agilent 3458A, \SI{10}{\volt} or \SI{100}{\volt} range), is applied to the gas-dielectric capacitor \oldrev{C} and converted to a current $I$, measured by the meter under calibration M. \newrev{The timer T provides a clock pulse to} all instruments, which are \newrev{controlled} by a personal computer PC. The current is computed from voltage samples with a simple finite-difference model \cite{Callegaro2007b}. 

\begin{figure}[htbp]
 	\includegraphics[width=\linewidth]{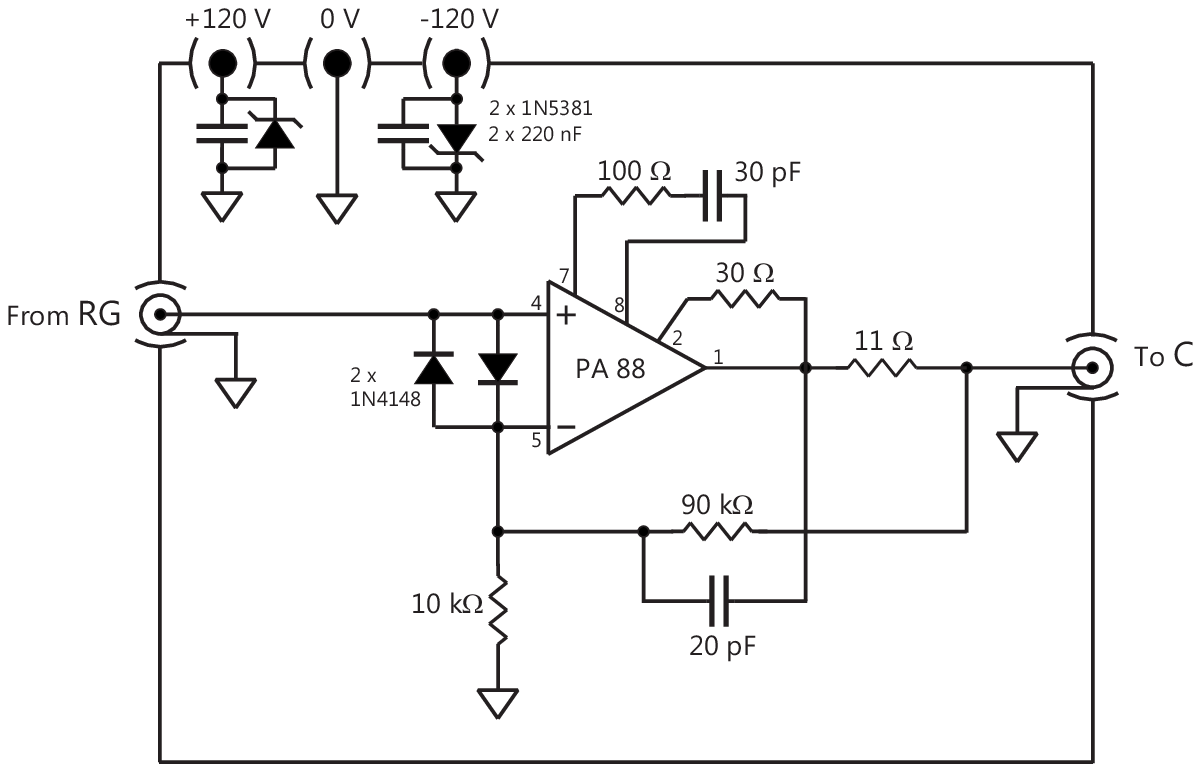}%
 	\caption{Range extender amplifier schematics, based on Apex mod. PA88 high-voltage power MOSFET integrated op amp, in a standard $\times10$ gain configuration. Diodes give input protection; resistor to pin 2 fixes the current limiter. Components between pins 7 and 8 are the datasheet recommended phase compensation network. Further compensation is given by the capacitor in the feedback network \cite{Franco2002}. \label{fig:amplischeme}}%
\end{figure}

The amplifier circuit is shown in Fig. \ref{fig:amplischeme}.

\section{Experimental}
\label{sec:experimental}
The quantity measured in all experiments is the \emph{calibration factor} $Q$ defined as
\begin{equation}
Q = \frac{I_\mathrm{r} - I_\mathrm{ro}}{I_\mathrm{g}}
\end{equation}
where $I_\mathrm{g}$ is the traceable current generated by the source, $I_\mathrm{r}$ is the current reading of M when $I_\mathrm{g}$ is applied, $I_\mathrm{ro}$ the current reading of M when zero current is applied.  

The range extender allows the generation of voltage ramps at $V_\mathrm{max}=\SI{100}{\volt}$ or higher\footnote{The op amp employed is rated for about \SI{450}{\volt} absolute maximum output.}, having properties very similar to the original ramp at $V_\mathrm{max}=\SI{10}{\volt}$ already described in \cite{Callegaro2007b}. \newrev{The} noise of E is specified as \SI{20}{\nano\volt \hertz^{-1/2}} at audio frequency. \newrev{Analyses} of samples acquired at \SI{1}{\hertz} rate when generating a \SI{100}{\volt} fixed output give a noise increase of \SI{40}{\micro\volt} rms when E is inserted. Although the input resistance of V \newrev{is reduced} to \SI{10}{\mega\ohm}, it does not constitute a significant load for E.

\newrev{The standard capacitors \oldrev{C} employed are General \oldrev{Radio} mod. 1404-B, \SI{100}{\pico\farad}, and 1404-C, \SI{1}{\nano\farad}, sealed dry nitrogen capacitors with Invar electrodes. The units are specified for voltages up to \SI{750}{\volt}, therefore the $V_\mathrm{max}$ increase is well tolerated. The units are calibrated as two terminal-pair standards \cite{CallegaroBook}. The applied voltage during calibration is \SI{15}{\volt} for \SI{100}{\pico\farad} and \SI{1.5}{\volt} for \SI{1}{\nano\farad}, and the frequency \SI{1}{\kilo\hertz}; very different conditions than those of the experiment. A determination of the voltage dependence of these units has not been performed, but it is known that in capacitors of similar construction is below \num{1E-7} in the \SIrange{10}{100}{\volt} rms range \cite{Hersh1963,Shields1965}. The frequency dependence is more of concern; in gas-dielectric capacitors it has been ascribed to surface effects \cite{Inglis1975,Giblin2010,Rietveld2012}. It has been found \cite{Giblin2010} that in 1404-C models the `ac-dc dependence', i.e. the capacitance deviation between the very low frequency employed in the experiment ($\approx \SI{10}{\milli\hertz}$) and \SI{1}{\kilo\hertz} (typically below \num{20E-6}, but with extremes up to \num{50E-6}) can be predicted from the measured dependence in the audio frequency range ($\SI{20}{\hertz}-\SI{1}{\kilo\hertz}$); for the item employed, the prediction gives an ac-dc dependence lower than \num{10E-6}.}

The calibration strategy and the related acquisiton, data processing software and uncertainty analysis described in \cite{Callegaro2007b, Willenberg2013} apply.

All experiments involve a top-class dc low current meter (Keithley mod. 6430A Sub-femtoamp remote sourcemeter) as the instrument under calibration M. The experiment are performed in an electromagnetically shielded room, \newrev{controlled} at \SI{23 \pm 0.5}{\celsius}.

An uncertainty budget for the calibration of M with the source including E, at \newrev{the} \SI{1}{\nano\ampere} level, is shown in Tab. \ref{tab:unccap1nA}.  \newrev{The uncertainty contribution for $C$ is conservative to include the voltage and frequency dependencies previously discussed}.

In order to validate the range extender, two different comparisons have been performed, where M is employed as a transfer standard. Since the instrument ranges are decadic (\SI{\pm 100}{\pico\ampere}, \SI{\pm 1}{\nano\ampere}) the effective nominal current of the comparisons is slightly lower (\SI{\pm 95}{\pico\ampere}, \SI{\pm 0.95}{\nano\ampere}) than the decadic value to avoid any possible systematic error caused by noise clipping.
\begin{table}[tb]
	\centering
	\caption{Calibration factor $Q$ at \SI{\pm 0.95}{\nano\ampere}: uncertainty budget  \label{tab:unccap1nA}}
	\begin{tabular}{lc}
		\addlinespace[5pt]
		\toprule
		Source & Uncertainty \\
			   & \SI{}{\micro\ampere \per \ampere} \\
		\midrule
		M resolution  									& $<1$ \\
		$C$												& 30 \\
		$v(t)$ sampling 									& 6 \\ 
		Timebase 										& 1 \\
		Current leakages									& 10 \\
		Reading noise									& 39 \\
		\midrule
		$Q$	 											& 51  \\
		\bottomrule
	\end{tabular}
\end{table}
\subsection{Comparison at \SI{100}{\pico\ampere}, including and excluding the range extender E}
The current source can generate a current of \SI{\pm 100}{\pico\ampere} in two different configurations, that is: by including the range extender E and employing a \SI{100}{\pico\farad} capacitance standard, or by excluding E and employing a \SI{1}{\nano\farad} capacitance standard. This allows a substitution comparison, \newrev{the results of which} are shown in Table~\ref{tab:comp100pA}; full agreement between the two configurations can be appreciated.

Since the configuration without E was validated in the context of an international intercomparison \cite{Willenberg2013}, the comparison is considered a proper validation of the configuration including E. 

\newrev{At the same time, the comparison gives a validation of the assumptions on the voltage and frequency dependencies of the capacitance standards employed.}
\begin{table}[tb]
	\centering
	\caption{Comparison at \SI{100}{\pico\ampere} \label{tab:comp100pA}}
	\begin{tabular}{ccccc}
		\toprule
		I & Extender & $V_\mathrm{max} $ & $C$ & $G$ \\
		\midrule
		\SI[retain-explicit-plus]{+95}{\pico\ampere}
		 		& no	& \SI{9.5}{\volt}	& \SI{1}{\nano\farad}  		& \num{1.000390(39)} \\
		\SI[retain-explicit-plus]{+95}{\pico\ampere} 
				& yes	& \SI{95}{\volt}	& \SI{100}{\pico\farad}  	& \num{1.000372(34)} \\
		\midrule
		\SI{-95}{\pico\ampere} 		& no	& \SI{9.5}{\volt}	& \SI{1}{\nano\farad} 		& \num{1.000337(37)} \\		
		\SI{-95}{\pico\ampere} 		& yes	& \SI{95}{\volt}	& \SI{100}{\pico\farad}  	& \num{1.000326(34)} \\
		\bottomrule
	\end{tabular}
\end{table}

\subsection{Comparison at \SI{1}{\nano\ampere}, versus Ohm's law method}
The comparison at \SI{\pm 1}{\nano\ampere} involves the capacitance-charging source (including E) and a standard calibration setup based on \newrev{the} Ohm's law method. The setup for Ohm's law method is based on a high-valued resistor $R$ and a calibrated dc voltage source $V$ and similar to that described in \cite{Callegaro2007, Galliana2010}.

In order to make the comparison meaningful and achieve a similar uncertainty for both methods, we chose $R=\SI{10}{\giga\ohm}$ and $V=\SI{\pm 10}{\volt}$. This required a special calibration of $R$ (standard calibrations are performed at higher voltages) which has been performed with the method described in \newrev{\cite{Henderson1987,Jarrett1997,Galliana2011}}. The corresponding uncertainty budget is shown in Table~\ref{tab:uncres10gohm}.  
\begin{table}[tb]
	\centering
	\caption{Calibration of $R=\SI{10}{\giga\ohm}$ at $V=\SI{10}{\volt}$: uncertainty budget \label{tab:uncres10gohm}}
	\begin{tabular}{lc}
		\addlinespace[5pt]
		\toprule
		Source & Uncertainty \\
			   & \SI{}{\micro\ohm \per \ohm} \\
		\midrule
		\SI{1}{\giga\ohm} standard				& 18 \\
		Bridge equilibrium						& 15 \\
		Leakage									& 3 \\
		Bridge ratio								& 11 \\
		Noise									& 19 \\
		\midrule
		$u(R)$									& 33   \\
		\bottomrule
	\end{tabular}
\end{table}

The uncertainty of the Ohm's law method at \SI{\pm 1}{\nano\ampere} has been evaluated and is given in Table~\ref{tab:uncohm1nA}. \oldrev{The contribution due to the voltage burden $V_\mathrm{B}$ of M appears dominant (Keithley 6430A specifications give $V_\mathrm{B}<$\SI{1}{\milli\volt}) for Keithley 6430A), but it can be much lower for other instruments (e.g., Keithley 6517B specifies $V_\mathrm{B}<$\SI{20}{\micro\volt}).}
\begin{table}[tb]
	\centering
	\caption{Ohm's law method, calibration factor $Q$ at \SI{\pm 0.95}{\nano\ampere}: uncertainty budget  \label{tab:uncohm1nA}}
	\begin{tabular}{lc}
		\addlinespace[5pt]
		\toprule
		Source & Uncertainty \\
			   & \SI{}{\micro\ampere \per \ampere} \\
		\midrule
		$R$ (Tab. \ref{tab:uncres10gohm})				& 33 \\
		$V$												& 3  \\
		M resolution										& \oldrev{$<1$}  \\
		M burden voltage									& 58 \\
		\oldrev{Noise}										& \oldrev{16} \\
		\midrule
		$u(Q)$ 											& 69  \\
		\bottomrule
	\end{tabular}
\end{table}
Results of the comparison at \SI{\pm 1}{\nano\ampere} are given in Table~\ref{tab:comp1nA}, showing good agreement between the two methods.
\begin{table}[t]
	\centering
	\caption{Comparison at \SI{1}{\nano\ampere} \label{tab:comp1nA}}
	\begin{tabular}{ccccc}
		\addlinespace[5pt]
		\toprule
		I & Method & $V$ or $V_\mathrm{max}$ & $R$ or $C$ & $Q$ \\
		\midrule
		\SI[retain-explicit-plus]{+0.95}{\nano\ampere} & 
			\oldrev{C} charging & \SI{95}{\volt}	& \SI{1}{\nano\farad} 	   					   & \num{1.000312(50)} \\		
		\SI[retain-explicit-plus]{+0.95}{\nano\ampere} 
			& Ohm's law & \SI[retain-explicit-plus]{+9.5}{\volt}	& \SI{10}{\giga\ohm}  		   & \num{1.000320(69)} \\		
		\midrule
		\SI{-0.95}{\nano\ampere} & \oldrev{C} charging & \SI{95}{\volt}	& \SI{1}{\nano\farad} 	   & \num{1.000265(50)} \\		
		\SI{-0.95}{\nano\ampere} & Ohm's law & \SI{-9.5}{\volt}	& \SI{10}{\giga\ohm}  		   & \num{1.000306(69)} \\		
		\bottomrule
	\end{tabular}
\end{table}
\section{Conclusions}
The capacitance-charging is a well-established method for accurate generation of low currents. Typical setups have a current limit in the \SI{100}{\pico\ampere} range. The range extender amplifier presented \newrev{here} allows an extension of available currents in the \SI{}{\nano\ampere} range; it has been employed as a plug-in component in the INRIM source, and can be adapted to other sources as well. The accuracy of the new setup allows a calibration of top-level meters with a relative uncertainty of \num{5E-5}; validation testing has been performed at two different current levels. 

A comparison at \SI{\pm 100}{\pico\ampere}, in the two configurations including and excluding the range extender, \newrev{was made} to test the extender on a point already validated through an international intercomparison. \newrev{The experiment adds confidence in the assumptions on the electrical behavior of the capacitance standards employed}.

The comparison at \SI{\pm 1}{\nano\ampere}, versus an implementation of Ohm's law method, showed good agreement with comparable uncertainties, so the methods are mutually validated and can both be employed. It must be however remarked that Ohm's law method \newrev{required} a special resistor calibration, and is very sensitive to the meter burden voltage.
\section*{Acknowledgments}
The authors are indebted with their INRIM colleagues Gian Carlo Bosco, for help during measurements, and Fabrizio Manta, who worked on the data acquisition software.

The work has been performed in the framework of Progetto Premiale MIUR-INRIM ``Nanotecnologie per la metrologia elettromagnetica''.
%
\bibliographystyle{IEEEtran}
\end{document}